\documentclass[11pt]{article}

\usepackage[utf8]{inputenc}
\usepackage[T1]{fontenc}
\usepackage{amsmath,amssymb}
\usepackage{graphicx}
\usepackage{booktabs}
\usepackage{hyperref}
\usepackage{url}
\usepackage{natbib}
\usepackage[margin=1in]{geometry}
\usepackage{xcolor}
\usepackage{listings}
\usepackage{enumitem}

\title{Review Beats Planning: Dual-Model Interaction Patterns\\for Code Synthesis}
\author{Jan Miller\\OPSWAT\\Hamburg, Germany\\\texttt{admin@jan-miller.de}}
\date{March 2026}

\begin{document}
\maketitle

% ============================================================================
\begin{abstract}
How should two language models interact to produce better code than either can alone? The conventional approach---a reasoning model \emph{plans}, a code specialist \emph{implements}---seems natural but fails: on HumanEval+, plan-then-code \textbf{degrades} performance by 2.4 percentage points versus the code specialist alone. We show that reversing the interaction changes everything. When the code specialist generates freely and the reasoning model \emph{reviews} instead of plans, the same two models on the same hardware achieve 90.2\% pass@1---exceeding GPT-4o (87.2\%) and O1~Preview (89.0\%)---on \textasciitilde\$2/hr of commodity GPU. Cross-benchmark validation across 542 problems (HumanEval+ and MBPP+) reveals a moderating variable: review effectiveness scales with \emph{specification richness}, yielding 4$\times$ more improvement on richly-specified problems (+9.8pp) than on lean ones (+2.3pp), while remaining net-positive in both cases. The practical implication is twofold: compose models by their cognitive strengths (reviewers review, coders code), and invest in specification quality to amplify the returns.
\end{abstract}

% ============================================================================
\section{Introduction}
\label{sec:intro}

Large language models have achieved impressive results on code generation benchmarks, with frontier models exceeding 87\% pass@1 on HumanEval+~\citep{evalplus}. However, these results typically rely on single large models (70B+ parameters) or proprietary APIs. For on-premise deployment, practitioners face a fundamental constraint: limited GPU memory forces the use of smaller, quantized models that individually fall short of state-of-the-art performance.

A natural response is to compose multiple specialized models into a pipeline. The conventional wisdom---inspired by software engineering practices and chain-of-thought reasoning~\citep{wei2022chain}---is to have a stronger reasoning model \emph{plan} the solution, then have a code specialist \emph{implement} that plan. We call this the ``plan-then-code'' pattern.

We test this assumption empirically and find it wrong. On HumanEval+, plan-then-code performs \emph{worse} than the raw code specialist, with the planner's guidance actively introducing errors in 15 of 164 problems. The key failure mode: the reasoning model suggests algorithms or identifies edge cases at a high level, but these suggestions sometimes conflict with the specific implementation constraints visible only in the code.

This finding motivates a reversal: instead of planning \emph{before} code generation, we apply the reasoning model \emph{after}---as a code reviewer. The ``review-then-fix'' pattern leverages the same models and the same compute budget, but changes the interaction direction. The result: a +10.4 percentage point improvement over baseline, achieving 90.2\% HumanEval+ with two 4-bit quantized models on commodity GPUs.

\paragraph{Contributions.}
\begin{enumerate}[leftmargin=*]
    \item \textbf{Empirical evidence that plan-then-code hurts.} We show that guidance from a reasoning model can degrade a code specialist's performance, with a detailed taxonomy of failure modes (\S\ref{sec:analysis}).
    \item \textbf{Review-then-fix as an alternative.} We propose and evaluate a dual-model pattern where the reasoning model reviews (not plans) and the code specialist fixes based on specific feedback (\S\ref{sec:review}).
    \item \textbf{Adversarial dual-generation.} We explore a pattern where both models generate solutions independently, which are then cross-validated and synthesized (\S\ref{sec:adversarial}).
    \item \textbf{Specification richness as a moderator.} Cross-benchmark validation on MBPP+ (378 problems) shows review effectiveness scales with specification detail: +9.8pp on rich specs vs +2.3pp on lean specs (\S\ref{sec:spec-richness}).
    \item \textbf{Practical on-premise setup.} All results use 4-bit AWQ quantized models on two A10G GPUs (\textasciitilde\$2/hr total), demonstrating that dual-model composition is a viable path to frontier-competitive performance.
\end{enumerate}

% ============================================================================
\section{Experimental Setup}
\label{sec:setup}

\subsection{Models}

\begin{table}[h]
\centering
\caption{Model configurations used in all experiments.}
\label{tab:models}
\begin{tabular}{llcccc}
\toprule
\textbf{Role} & \textbf{Model} & \textbf{Params} & \textbf{Quant.} & \textbf{GPU} & \textbf{VRAM} \\
\midrule
Code specialist & Qwen2.5-Coder-14B-Instruct & 14B & AWQ 4-bit & A10G \#1 & \textasciitilde8\,GB \\
Reasoning generalist & Qwen3-32B & 32B & AWQ 4-bit & A10G \#2 & \textasciitilde18\,GB \\
\bottomrule
\end{tabular}
\end{table}

Both models are served via vLLM~\citep{vllm} using the OpenAI Chat Completions API format. We use greedy decoding (temperature\,=\,0) for deterministic, reproducible results.

\paragraph{Qwen3 thinking mode.} Qwen3-32B supports an explicit reasoning mode that emits internal reasoning tokens before the response. For raw benchmarking, we disable thinking to avoid output pollution. For pipeline configurations where the model performs analysis or review, we enable thinking to leverage the model's reasoning capability.

\subsection{Benchmark}

We use two benchmarks from the EvalPlus framework~\citep{evalplus}:
\begin{itemize}[leftmargin=*]
    \item \textbf{HumanEval+}: 164 Python problems with rich specifications---function signatures, detailed docstrings, type annotations, and doctest examples. Augmented with 80$\times$ more tests than the original HumanEval~\citep{humaneval}.
    \item \textbf{MBPP+}: 378 Python problems with lean specifications---typically a one-line description and a single assertion. Augmented test suite from EvalPlus.
\end{itemize}
This pair provides a natural experiment in specification richness: same evaluation framework, same models, but dramatically different specification detail. We report pass@1 on the augmented suites (``HumanEval+'' and ``MBPP+'') with greedy decoding.

\subsection{Code Extraction and Validation}

Model outputs are post-processed to extract function bodies from potentially verbose responses, handling markdown fence stripping, repeated signature removal, indentation normalization, and trailing text truncation. Each extraction step includes compile-time verification.

Pipeline variants with retry include an \emph{eval-retry loop}: generated code is compiled, executed, and tested against docstring examples visible in the prompt. On failure, the error message is fed back for correction (up to 3 retries). Importantly, this uses only tests visible in the prompt---not the hidden EvalPlus augmented tests---preventing any form of test leakage.

% ============================================================================
\section{Pipeline Architectures}
\label{sec:pipelines}

\begin{figure}[h]
\centering
\includegraphics[width=0.85\textwidth]{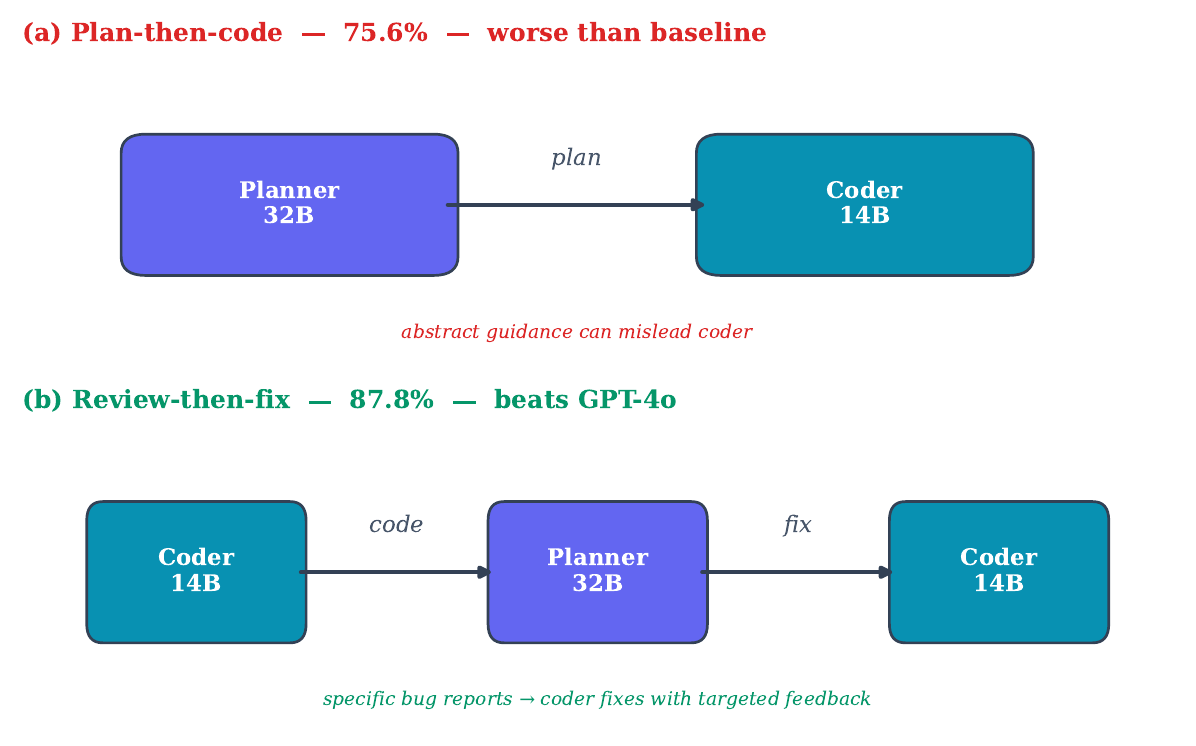}
\caption{Two dual-model interaction patterns. (a)~Plan-then-code: the reasoning model plans upfront, but its abstract guidance can mislead the coder (75.6\%). (b)~Review-then-fix: the coder generates freely, the reasoning model reviews for bugs, and the coder fixes---achieving 87.8\% without retry.}
\label{fig:architecture}
\end{figure}

Figure~\ref{fig:architecture} contrasts the two primary interaction patterns.

\subsection{Plan-Then-Code}
\label{sec:plan}

The planner receives the function signature and docstring, produces a natural-language analysis (algorithm, edge cases, complexity), and the coder implements with this analysis as context. This is the conventional ``chain-of-thought delegation'' pattern.

\subsection{Review-Then-Fix}
\label{sec:review}

The key insight: instead of telling the code specialist \emph{how} to solve the problem (which can mislead), let it solve the problem naturally, then use the reasoning model to \emph{find bugs}. The code specialist then receives specific, actionable feedback (``line X does Y, but should do Z'') rather than abstract algorithmic guidance.

The pipeline proceeds as:
\begin{enumerate}
    \item Coder generates a solution (identical prompt to raw baseline)
    \item Planner reviews the solution against the specification
    \item If bugs found, coder receives the review and fixes (single pass)
\end{enumerate}

This core pipeline uses exactly 2--3 LLM calls per problem: one generation, one review, and optionally one fix. No test execution, no retry loop---a fair comparison against single-shot leaderboard entries.

We additionally evaluate a variant \textbf{review-then-fix + retry} that adds an eval-retry loop: after the review-fix pass, the solution is compiled and tested against docstring examples visible in the prompt. On failure, the error is fed back for correction (up to 3 retries). This variant is reported separately to isolate the contribution of review from the contribution of test feedback.

Additional context provided to the coder includes an import inventory (explicitly stating allowed imports).

\subsection{Adversarial Dual-Generation}
\label{sec:adversarial}

This architecture eliminates ``guidance contamination'' entirely:
\begin{enumerate}
    \item Both models generate solutions independently (in parallel)
    \item Cross-validate both against docstring tests
    \item If exactly one passes, use it; if both pass, adversarial review selects the better one
    \item If neither passes, the planner reviews both; the coder synthesizes a third solution informed by both failures and their reviews
\end{enumerate}

% ============================================================================
\section{Results}
\label{sec:results}

\subsection{Main Results}

\begin{table}[h]
\centering
\caption{Pass@1 results on HumanEval and HumanEval+ (greedy decoding, 164 problems).}
\label{tab:results}
\begin{tabular}{lccc}
\toprule
\textbf{Configuration} & \textbf{HumanEval} & \textbf{HumanEval+} & \textbf{$\Delta$ vs baseline} \\
\midrule
raw-coder (14B AWQ) & 81.1\% & 78.0\% & --- \\
raw-planner (32B AWQ) & 89.0\% & 84.1\% & +6.1pp \\
\midrule
Plan-then-code & 80.5\% & 75.6\% & $-$2.4pp \\
\midrule
Review-then-fix (no retry) & 89.6\% & 87.8\% & +9.8pp \\
Review-then-fix (+retry) & 93.3\% & 90.2\% & +12.2pp \\
Adversarial debate & 91.5\% & 86.6\% & +8.6pp \\
\bottomrule
\end{tabular}
\end{table}

\begin{figure}[h]
\centering
\includegraphics[width=0.95\textwidth]{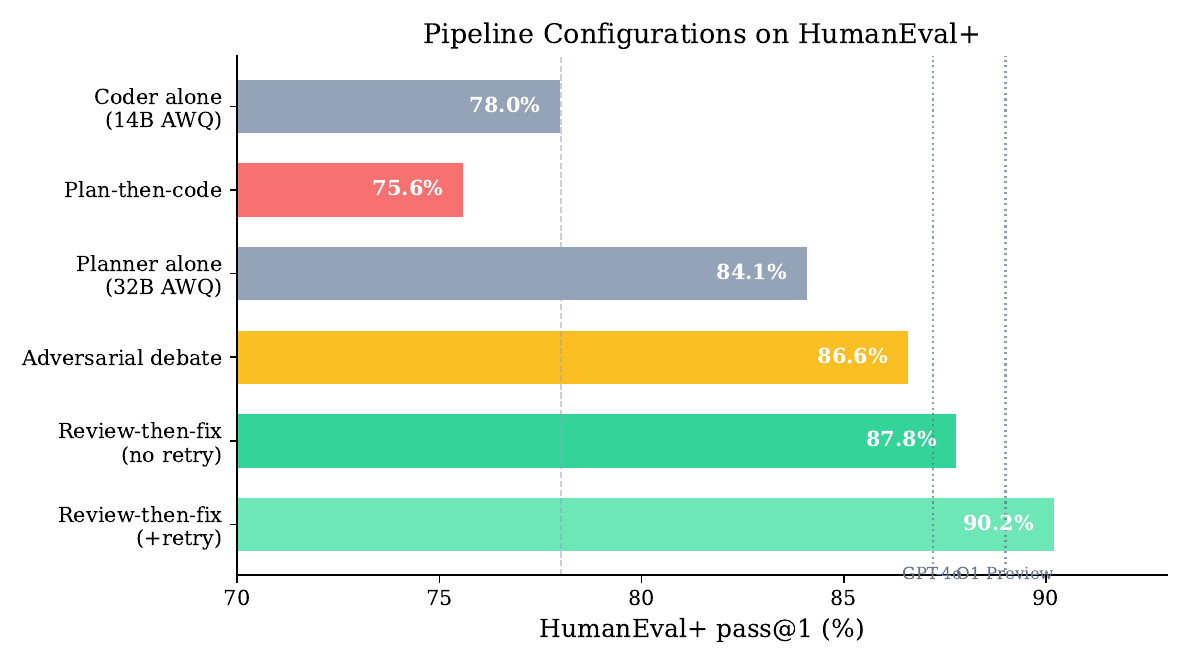}
\caption{Pipeline configurations on HumanEval+. Plan-then-code (red) performs \emph{worse} than the raw coder baseline. Review-then-fix (green) exceeds both GPT-4o and O1~Preview. Dashed line: coder baseline; dotted lines: proprietary model reference points.}
\label{fig:main_results}
\end{figure}

\subsection{Cross-Benchmark Validation: Specification Richness}
\label{sec:spec-richness}

To understand \emph{when} review helps, we replicated our key comparison on MBPP+ (378 problems)---a benchmark with substantially leaner specifications than HumanEval+. Where HumanEval+ provides function signatures, detailed docstrings, type annotations, and doctest examples, MBPP+ typically offers a one-line natural language description and a single assertion.

\begin{table}[h]
\centering
\caption{Review effectiveness scales with specification richness. Comparison of review-then-fix (no retry) across benchmarks.}
\label{tab:spec-richness}
\begin{tabular}{lcccc}
\toprule
& \multicolumn{2}{c}{\textbf{HumanEval+ (rich specs)}} & \multicolumn{2}{c}{\textbf{MBPP+ (lean specs)}} \\
\cmidrule(lr){2-3} \cmidrule(lr){4-5}
\textbf{Configuration} & \textbf{Pass@1} & \textbf{$\Delta$} & \textbf{Pass@1} & \textbf{$\Delta$} \\
\midrule
raw-coder (baseline) & 78.0\% & --- & 67.5\% & --- \\
Review-then-fix & 87.8\% & +9.8pp & 69.8\% & +2.3pp \\
Enriched review & 88.4\% & +10.4pp & 69.8\% & +2.3pp \\
\bottomrule
\end{tabular}
\end{table}

The same review pipeline that yields +9.8pp on HumanEval+ yields only +2.3pp on MBPP+---a 4$\times$ reduction in review impact. We attribute this to the reviewer's ability to assess correctness: with rich specifications (examples, types, edge cases), the reviewer has clear criteria against which to judge the code. With lean specifications, the reviewer must infer intent from a brief description, increasing the chance of both missed bugs and false positives.

\begin{figure}[h]
\centering
\includegraphics[width=0.95\textwidth]{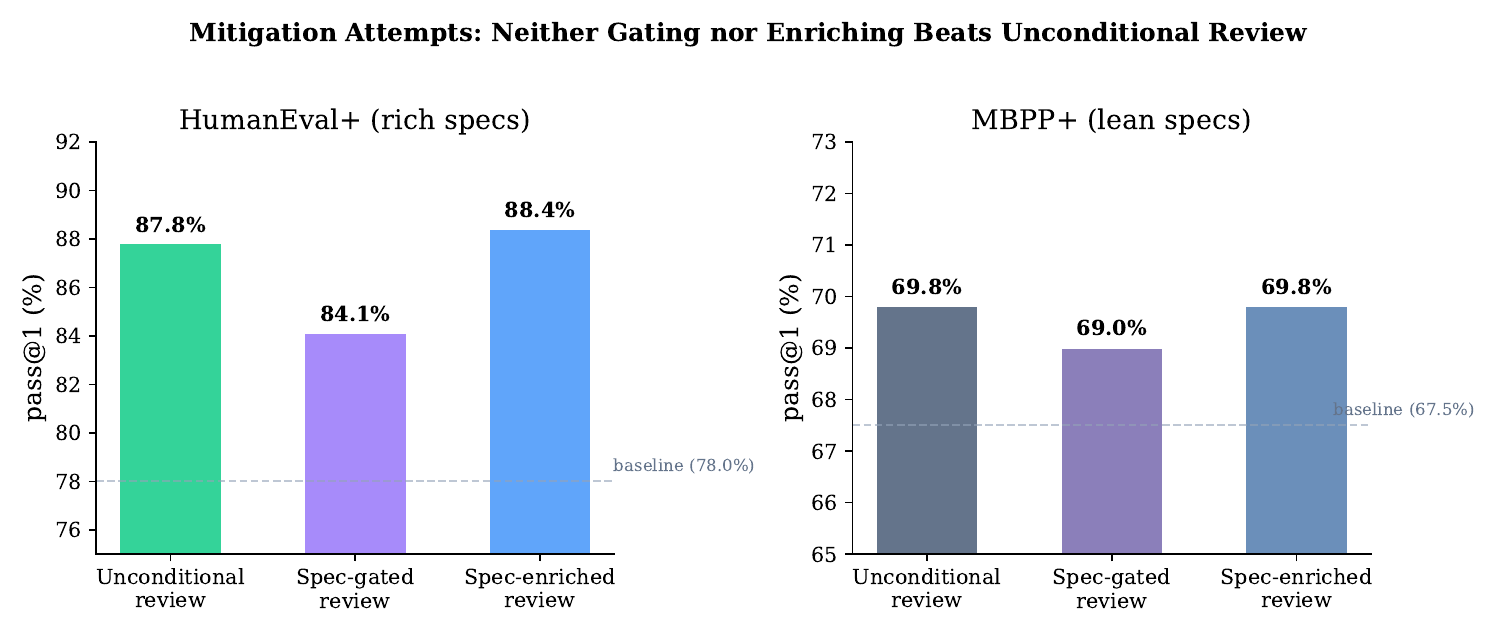}
\caption{Mitigation attempts for the spec-richness gap. Left: HumanEval+ (rich specs). Right: MBPP+ (lean specs). Neither spec-gated review (purple) nor spec-enriched review (blue) outperforms unconditional review (green). Dashed line: coder-alone baseline.}
\label{fig:mitigations}
\end{figure}

Crucially, review is still \emph{net-positive} on both benchmarks. We tested two mitigations for the lean-spec gap:
\begin{itemize}[leftmargin=*]
    \item \textbf{Spec-gated review:} Selectively skip review for lean specifications. Performed \emph{worse} than unconditional review (84.1\% HumanEval+, 69.0\% MBPP+)---the modest false-positive rate on lean specs is offset by the genuine bugs review catches.
    \item \textbf{Spec-enriched review:} Use the reasoning model to generate additional examples, types, and edge cases before review. Marginal gain on HumanEval+ (+0.6pp) and no gain on MBPP+ (69.8\% vs 69.8\%), at the cost of an additional LLM call per problem.
\end{itemize}
Neither mitigation closes the gap (Figure~\ref{fig:mitigations}). The practical implication: unconditional review is the right default, specification richness is a fundamental moderator that cannot be cheaply synthesized, and investment in upstream specification quality is the most effective lever for improving review outcomes.

\begin{figure}[h]
\centering
\includegraphics[width=0.95\textwidth]{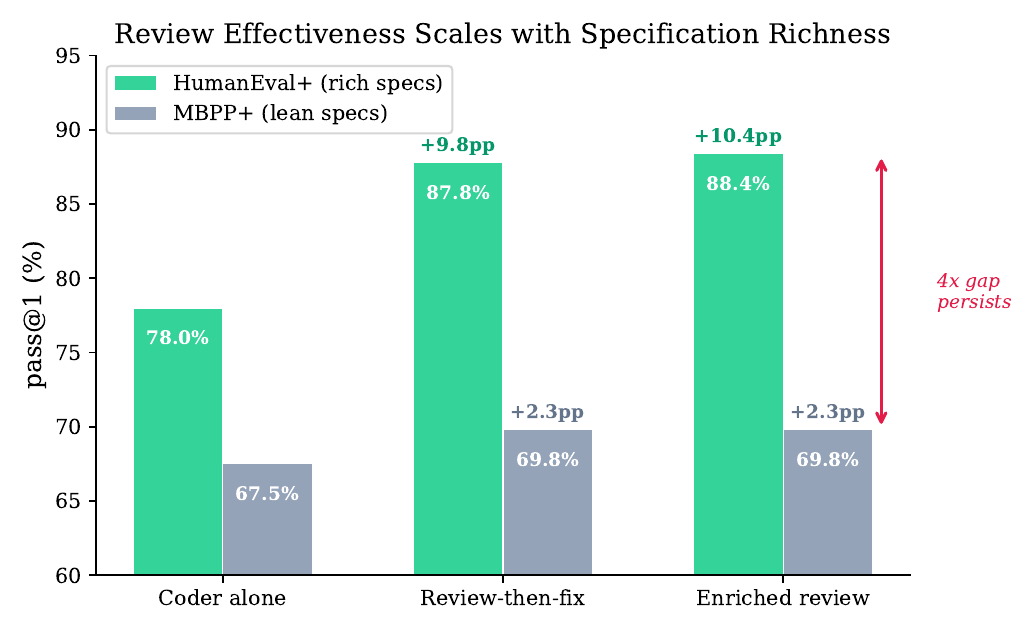}
\caption{Review effectiveness scales with specification richness. The 4$\times$ gap between rich specs (HumanEval+, +9.8pp) and lean specs (MBPP+, +2.3pp) persists even with spec enrichment. Green: HumanEval+ (rich); gray: MBPP+ (lean).}
\label{fig:spec_richness}
\end{figure}

\subsection{Leaderboard Context}

\begin{table}[h]
\centering
\caption{Comparison with published HumanEval+ results from the EvalPlus leaderboard.}
\label{tab:leaderboard}
\begin{tabular}{lcc}
\toprule
\textbf{Model} & \textbf{HumanEval+} & \textbf{Notes} \\
\midrule
O1 Preview & 89.0\% & Proprietary reasoning model \\
\textbf{Our pipeline (14B+32B AWQ)} & \textbf{90.2\%} & 2$\times$ A10G, \textasciitilde\$2/hr \\
Qwen2.5-Coder-32B (FP16) & 87.2\% & Single model, full precision \\
GPT-4o & 87.2\% & Proprietary \\
DeepSeek-V3 & 86.6\% & Proprietary \\
\bottomrule
\end{tabular}
\vspace{0.5em}

{\small Full review-then-fix run on all 164 problems (not merged).}
\end{table}

\subsection{Failure Mode Analysis}
\label{sec:analysis}

\begin{figure}[h]
\centering
\includegraphics[width=0.5\textwidth]{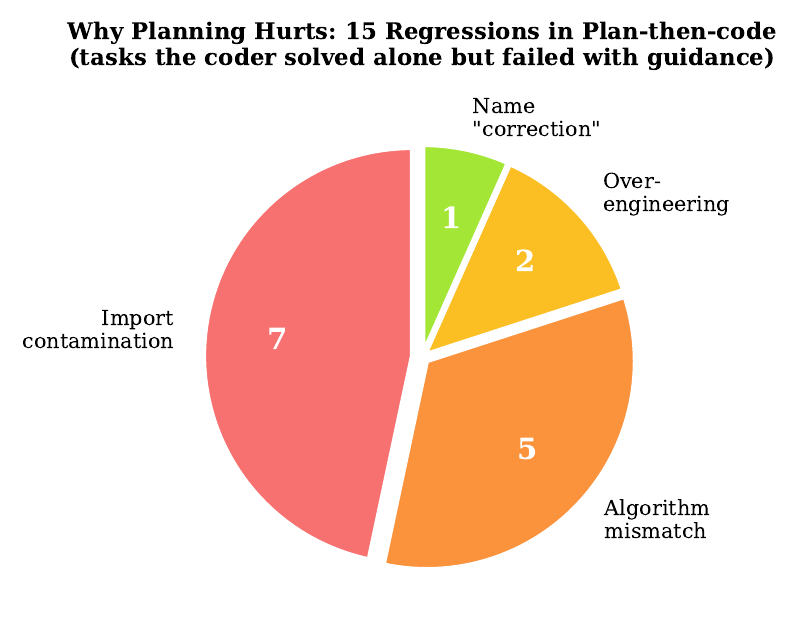}
\caption{Breakdown of 15 regressions in plan-then-code: tasks the coder solved alone but failed with planner guidance. Import contamination dominates---the planner suggests libraries the coder then uses without importing.}
\label{fig:failure_modes}
\end{figure}

Of 164 problems, plan-then-code produced 15 regressions where the pipeline broke tasks the raw coder solved, against only 14 improvements. We categorize the regressions (Figure~\ref{fig:failure_modes}):

\begin{itemize}[leftmargin=*]
    \item \textbf{Missing imports (7):} Planner suggests using \texttt{Counter}, \texttt{math.factorial}, etc. Coder follows the suggestion but omits the import.
    \item \textbf{Variable name ``correction'' (1):} Problem uses intentional misspelling (\texttt{delimeter}). Planner ``fixes'' it to \texttt{delimiter}, causing signature mismatch.
    \item \textbf{Algorithm mismatch (5):} Planner's high-level approach is correct but coder introduces subtle bugs during translation.
    \item \textbf{Over-engineering (2):} Planner identifies edge cases that don't exist, adding unnecessary complexity.
\end{itemize}

\subsection{Why Review Beats Planning}

We hypothesize three mechanisms:
\begin{enumerate}
    \item \textbf{Specificity of feedback.} Planning produces abstract suggestions; review produces specific bug reports. Specific feedback is easier to act on.
    \item \textbf{Preserved code autonomy.} The coder first solves in its natural style, then receives corrections---its internal consistency is preserved.
    \item \textbf{Asymmetric strengths.} The 32B generalist excels at \emph{finding bugs} (reasoning); the 14B specialist excels at \emph{generating code}. Review-then-fix aligns each model with its strength.
\end{enumerate}

% ============================================================================
\section{Related Work}
\label{sec:related}

\paragraph{Self-repair.} \citet{chen2023selfdebugging} and \citet{olausson2024selfrepair} study single-model code repair using execution feedback. Our work differs by using a \emph{separate} reasoning model for review, introducing genuine model diversity.

\paragraph{Multi-agent code generation.} ChatDev~\citep{chatdev} and MetaGPT~\citep{metagpt} use multi-agent systems for software development, but typically role-play with the same base model. We use genuinely different models (specialist vs generalist) and show that interaction direction matters more than role assignment.

\paragraph{LLM composition.} Speculative decoding, routing, and mixture-of-experts combine models for efficiency. Our work composes models for \emph{quality}---achieving results neither model reaches alone.

% ============================================================================
\section{Discussion and Limitations}
\label{sec:discussion}

\textbf{Benchmark scope.} HumanEval+ tests single-function problems, not real-world multi-file development.

\textbf{Eval-retry uses visible tests.} The docstring test validation is a lightweight correctness signal, not a substitute for comprehensive testing.

\textbf{Cost trade-off.} Review-then-fix requires 2--7 LLM calls per problem vs 1 for single-model generation.

\textbf{Model specificity.} Results are specific to Qwen2.5-Coder-14B + Qwen3-32B; other combinations may differ.

\textbf{Specification enrichment does not close the gap.} Since review effectiveness scales with specification richness, a natural mitigation is using the reasoning model to \emph{enrich} lean specifications before review---generating examples, types, and edge cases from brief descriptions. We tested this (``enriched review'': coder generates, planner enriches spec, planner reviews against enriched spec, coder fixes). On HumanEval+ (already rich specs), enrichment yields a marginal +0.6pp (88.4\% vs 87.8\% unconditional review). On MBPP+ (lean specs), enrichment shows no improvement (69.8\% vs 69.8\%). The auto-generated enrichment does not substitute for genuine specification detail, suggesting that the specification richness gap is a fundamental property of the problem, not an artifact of missing context that can be synthesized.

% ============================================================================
\section{Conclusion}
\label{sec:conclusion}

We demonstrate that dual-model composition can achieve frontier-competitive code synthesis (90.2\% HumanEval+) using two 4-bit quantized open models on commodity GPUs. The critical finding is that \emph{interaction pattern selection}---specifically, using the reasoning model for code review rather than upfront planning---is the primary driver of quality improvement. Plan-then-code degrades performance; review-then-fix dramatically improves it. Cross-benchmark validation on MBPP+ reveals a second finding: review effectiveness scales with \emph{specification richness}, yielding 4$\times$ more improvement on richly-specified problems (+9.8pp on HumanEval+) than on lean ones (+2.3pp on MBPP+), while remaining net-positive in both cases. This suggests that in dual-model systems, models should be composed according to their cognitive strengths rather than following the human metaphor of plan-then-implement, and that investment in specification quality amplifies the returns of code review.

% ============================================================================
\bibliographystyle{plainnat}
\bibliography{references}

@inproceedings{evalplus,
  title={Is Your Code Generated by {ChatGPT} Really Correct? Rigorous Evaluation of Large Language Models for Code Generation},
  author={Liu, Jiawei and Xia, Chunqiu Steven and Wang, Yuyao and Zhang, Lingming},
  booktitle={NeurIPS},
  year={2023}
}

@inproceedings{wei2022chain,
  title={Chain-of-Thought Prompting Elicits Reasoning in Large Language Models},
  author={Wei, Jason and Wang, Xuezhi and Schuurmans, Dale and Bosma, Maarten and Ichter, Brian and Xia, Fei and Chi, Ed and Le, Quoc and Zhou, Denny},
  booktitle={NeurIPS},
  year={2022}
}

@inproceedings{vllm,
  title={Efficient Memory Management for Large Language Model Serving with {PagedAttention}},
  author={Kwon, Woosuk and Li, Zhuohan and Zhuang, Siyuan and Ying, Ying and Lian, Ying and Cho, Kangmin and Zhuang, Hao and Gonzalez, Joseph E and Stoica, Ion and Xing, Eric P},
  booktitle={SOSP},
  year={2023}
}

@article{humaneval,
  title={Evaluating Large Language Models Trained on Code},
  author={Chen, Mark and Tworek, Jerry and Jun, Heewoo and Yuan, Qiming and de Oliveira Pinto, Henrique Pond{\'e} and Kaplan, Jared and Edwards, Harri and Burda, Yuri and Joseph, Nicholas and Brockman, Greg and others},
  journal={arXiv preprint arXiv:2107.03374},
  year={2021}
}

@article{chen2023selfdebugging,
  title={Teaching Large Language Models to Self-Debug},
  author={Chen, Xinyun and Lin, Maxwell and Sch{\"a}rli, Nathanael and Zhou, Denny},
  journal={arXiv preprint arXiv:2304.05128},
  year={2023}
}

@article{olausson2024selfrepair,
  title={Is Self-Repair a Silver Bullet for Code Generation?},
  author={Olausson, Theo X and Inala, Jeevana Priya and Wang, Chenglong and Gao, Jianfeng and Solar-Lezama, Armando},
  journal={ICLR},
  year={2024}
}

@inproceedings{chatdev,
  title={Communicative Agents for Software Development},
  author={Qian, Chen and Cong, Xin and Yang, Cheng and Chen, Weize and Su, Yusheng and Xu, Juyuan and Liu, Zhiyuan and Sun, Maosong},
  booktitle={ACL},
  year={2024}
}

@inproceedings{metagpt,
  title={{MetaGPT}: Meta Programming for A Multi-Agent Collaborative Framework},
  author={Hong, Sirui and Zhuge, Mingchen and Chen, Jonathan and Zheng, Xiawu and Cheng, Yuheng and Zhang, Ceyao and Wang, Jinlin and Wang, Zili and Yau, Steven Ka Shing and Lin, Zijuan and others},
  booktitle={ICLR},
  year={2024}
}

\end{document}